%% file: acl_latex.tex
\pdfoutput=1
\documentclass[11pt]{article}

\usepackage[final]{acl}
\usepackage{times}
\usepackage{latexsym}
\usepackage[T1]{fontenc}
\usepackage[utf8]{inputenc}
\usepackage{microtype}
\usepackage{inconsolata} 
\usepackage{subcaption}
\usepackage{graphicx}
\usepackage{amsmath}
\usepackage{amssymb}
\usepackage{tabularx}
\usepackage{booktabs} 
\usepackage[table]{xcolor}
\usepackage{multirow} 
\usepackage{algorithm}   
\usepackage{algorithmic}  
\usepackage{caption}      
\usepackage{xspace}
\usepackage{enumitem}
\usepackage{placeins}

\newcommand{\boldstart}[1]{\noindent \textbf{#1}}

\newcommand{\systemname}[0]{\textmd{DFSU}\xspace}

\title{Data-Free Privacy-Preserving for LLMs via Model Inversion and Selective Unlearning}

\author{%
  Xinjie Zhou$^{1,2}$,\,
  Zhihui Yang$^{1,2,4,5}$\thanks{Corresponding author.},\,
  Lechao Cheng$^{3}$,\,
  Sai Wu$^{2,6,7}$,\,
  Gang Chen$^{2}$\\[2mm]
  $^{1}$School of Software Technology, Zhejiang University\\
  $^{2}$Zhejiang University\\
  $^{3}$Hefei University of Technology\\
  $^{4}$Institute of Fundamental and Transdisciplinary Research, Zhejiang University\\
  $^{5}$Hangzhou High-Tech Zone (Binjiang) Institute of Blockchain and Data Security\\
  $^{6}$The State Key Laboratory of Blockchain and Data Security\\
  $^{7}$Zhejiang Key Laboratory of Big Data Intelligent Computing\\[2pt]
  {\tt \{xinjiezhou,zhyangcs, wusai, cg\}@zju.edu.cn,\ \ chenglc@hfut.edu.cn}\\[2pt]
  \vadjust{\vspace{2pt}}
}

\begin{document}
\maketitle
\input{sections/0abstract}
\input{sections/1introduction}
\input{sections/2relatedwork}

\input{sections/3method}

\input{sections/4setup}

\input{sections/5evaluation}

\input{sections/6conclusion}
\input{sections/7limitations}
\bibliography{custom}
\clearpage
\input{sections/8appendix}
\end{document}

%% file: sections/0abstract.tex
\begin{abstract}
Large language models (LLMs) exhibit powerful capabilities but risk memorizing sensitive personally identifiable information (PII) from their training data, posing significant privacy concerns. While machine unlearning techniques aim to remove such data, they predominantly depend on access to the training data. This requirement is often impractical, as training data in real-world deployments is commonly proprietary or inaccessible. To address this limitation, we propose Data-Free Selective Unlearning (\systemname), a novel privacy-preserving framework that removes sensitive PII from an LLM without requiring its training data. Our approach first synthesizes pseudo-PII through language model inversion, then constructs token-level privacy masks for these synthetic samples, and finally performs token-level selective unlearning via a contrastive mask loss within a low-rank adaptation (LoRA) subspace. Extensive experiments on the AI4Privacy PII-Masking dataset using Pythia models demonstrate that our method effectively removes target PII while maintaining model utility. 

\end{abstract}

%% file: sections/1introduction.tex
\section{Introduction}

Recent advances in large language models (LLMs) have transformed a wide range of applications, but they also raise acute privacy concerns: internet-scale pre-training corpora inevitably contain personally identifiable information (PII)~\cite{carlini2021extracting,carlini2023quantifying}, and LLMs can inadvertently memorize and later reproduce such content (e.g., addresses or medical records), creating substantial legal, ethical, and safety risks in deployment. 

\begin{figure}[t]        
    \centering
    \includegraphics[width=\columnwidth]{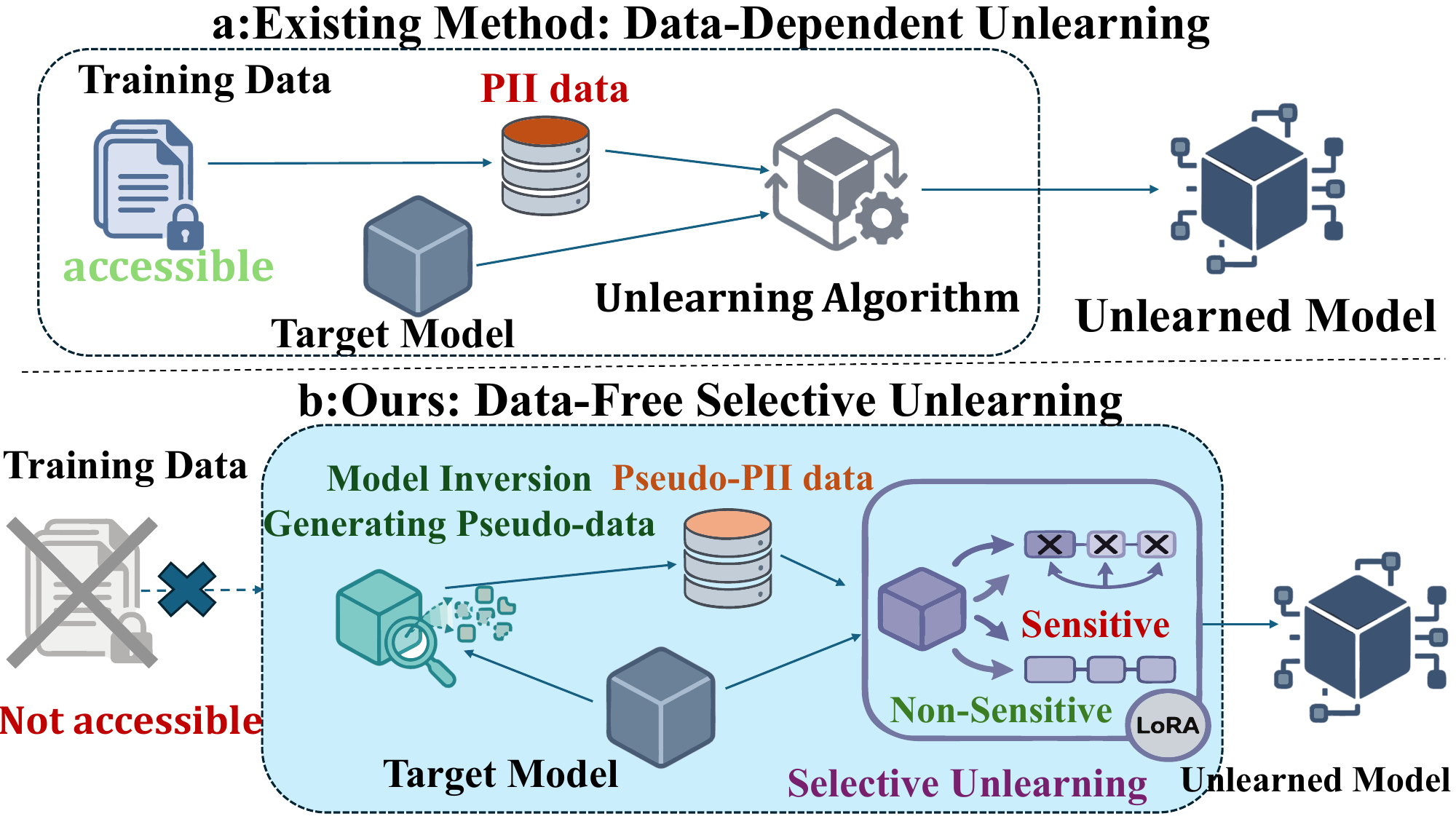}
    \caption{A comparison of (a) data-dependent unlearning and (b) data-free selective unlearning.}
    \label{fig:comparison}
    \vspace{-5mm}
\end{figure}

To mitigate these risks, \emph{machine unlearning}~\cite{bourtoule2021machine} has emerged as a key direction. Existing approaches largely fall into two paradigms: \textbf{exact unlearning}~\cite{chowdhury2025towards,muresanu2025fast}, which retrains from scratch but is computationally prohibitive for LLMs, and \textbf{approximate unlearning}~\cite{yao2024machine,chang2024do}, which updates model parameters to forget specific data. Despite progress, a fundamental limitation persists: most methods remain intrinsically \emph{data-dependent}~\cite{cao2015towards,muresanu2025fast}. As illustrated in Fig~\ref{fig:comparison} (a), representative techniques such as Gradient Ascent (GA)~\cite{jang2023knowledge,yao2023large} and Negative Preference Optimization (NPO)~\cite{zhang2024negative} require access to the original training corpus or an explicit ``forget set'' to compute unlearning gradients~\cite{liu2024rethinking}. In practice, this assumption often fails: practitioners may only have access to model weights, while the training data can be proprietary~\cite{touvron2023llama}, legally restricted under ``Right to be Forgotten'' regulations~\cite{liu2024rethinking}, or simply unrecoverable at scale~\cite{gao2020pile}. Consequently, current unlearning methods can become inapplicable precisely in the settings where post-hoc privacy remediation is most needed.

Motivated by the cognitive phenomenon that specific memories can be attenuated by suppressing internal representations without re-exposure to sensitive contents, we study \emph{data-free selective unlearning} (Fig.~\ref{fig:comparison}b): removing memorized PII from a pre-trained LLM \emph{post hoc}, using only model parameters and without accessing to the original pre-training corpus. This setting is challenging because selective unlearning requires \emph{localized} interventions---privacy-relevant behaviors must be suppressed while the model's general linguistic and reasoning capabilities are preserved. Without explicit data supervision, the optimization signal is weakly constrained, and naive updates diffuse across entangled representations, leading to either incomplete privacy removal or unnecessary utility degradation.

A key practical observation is that defenders often know the \emph{type} of information to be forgotten (e.g., IP addresses, device identifiers) even when the exact training instances are unavailable. We leverage this prior as a directional cue and propose to synthesize an effective surrogate ``forget set'' via model inversion, repurposing inversion attacks as a defensive tool. Building on this insight, we introduce \systemname, a data-free privacy-preserving framework that removes sensitive PII from an LLM without accessing its original training data. \systemname follows a three-stage pipeline: (i) we train a logit-based inversion model to capture memorized PII patterns from a target LLM; (ii) we generate pseudo-PII samples and annotate them via few-shot prompting; and (iii) we perform parameter-efficient selective unlearning in a LoRA adaptation space, using a contrastive masking objective to suppress identified sensitive tokens while anchoring surrounding context to preserve utility.

We evaluate \systemname on both generative (WikiText-103)~\cite{merity2016pointer} and reasoning/classification (MNLI)~\cite{williams-etal-2018-broad} tasks using pretrained Pythia models (160M/410M/1.4B)~\cite{biderman2023pythia} and sensitive data from the AI4Privacy dataset~\cite{ai4privacy2024pii}. Across scales and scenarios, \systemname consistently approaches the privacy--utility balance achieved by an oracle that unlearns with access to the original training data, demonstrating a practical path to post-hoc privacy remediation in data-restricted deployments. Our contributions are summarized as follows:

\begin{itemize}[leftmargin=*]
    \item We formalize the problem of data-free selective unlearning, addressing the critical challenge of performing privacy preservation when the original training data is inaccessible.
    \item We propose~\systemname, a novel three-stage pipeline that integrates model inversion, pseudo-PII synthesis, and selective token-level unlearning to remove memorized PII from pretrained LLMs without accessing their training data.
    \item Through comprehensive experiments on both generative and classification tasks, we show that \systemname achieves a privacy-utility trade-off competitive with Oracle-based unlearning.
\end{itemize}

%% file: sections/2relatedwork.tex
\section{Related Work}\label{sec:related_work}

\boldstart{Privacy Risks in LLMs.}
LLMs behave as probabilistic databases and can exhibit strong memorization of their training corpora~\citep{carlini2021extracting}. This risk scales with model capacity: larger models disproportionately retain long-tail content, which often includes sensitive PII~\citep{carlini2023quantifying}. Such memorization is exploitable via extraction attacks (e.g., prefix probing) and membership inference, enabling adversaries to recover private records~\citep{nasr2023scalable}. While training-time defenses such as DP-SGD provide formal guarantees~\citep{abadi2016deep}, they typically degrade utility and are not retroactive—once leakage is found in a deployed model, they cannot remediate it. This gap motivates post-hoc unlearning mechanisms for privacy mitigation after pre-training.

\boldstart{Machine Unlearning.} Most post-hoc unlearning methods are intrinsically data-dependent, requiring access to ground-truth sensitive examples. Gradient-ascent (GA) approaches~\citep{jang2023knowledge} maximize loss on private samples but can induce catastrophic collapse, degrading general language competence alongside the targeted facts~\citep{yuan2025closer,xing2025knowledge}. Negative Preference Optimization (NPO)~\citep{zhang2024negative} mitigates instability by anchoring updates to a reference distribution, yet still assumes a precisely specified forget set. Related model-editing work frames unlearning as localized knowledge suppression: for instance, Private Memorization Editing (PME)~\citep{ruzzetti2025private} first detects memorized PII via extraction and then edits feed-forward layers to reduce its emission. These lines of work share a critical prerequisite---access to training data or original sensitive samples to identify, localize, and suppress memorized PII~\citep{liu2024rethinking,ruzzetti2025private}. Such data is proprietary, legally restricted, or unavailable, rendering these methods impractical. By design, \systemname targets this data-free regime and performs selective privacy remediation using only model weights and defender-specified priors.

\boldstart{Model Inversion.}
Model inversion has been traditionally studied as an adversarial threat, aiming to reconstruct training inputs from model representations or outputs. Methods such as Vec2Text~\citep{morris2023text} and logit-based inversion~\citep{zhang2022text} recover textual inputs via optimization or learned decoders. While recent work primarily focuses on defending against such attacks~\citep{chen2025algen}, we propose a paradigm shift by leveraging inversion for defensive purposes. By treating a model's own memorized logits as a generative prior, our~\systemname synthesizes privacy-relevant pseudo-samples to bridge the data-free gap. Crucially, instead of retraining on noisy synthetic data, we integrate inversion with token-level selective masking to suppress target PII without requiring access to the original training data.

%% file: sections/3method.tex
\section{Methodology}
\label{sec:method}

\begin{figure*}[t]
    \centering
    \includegraphics[width=0.9\textwidth]{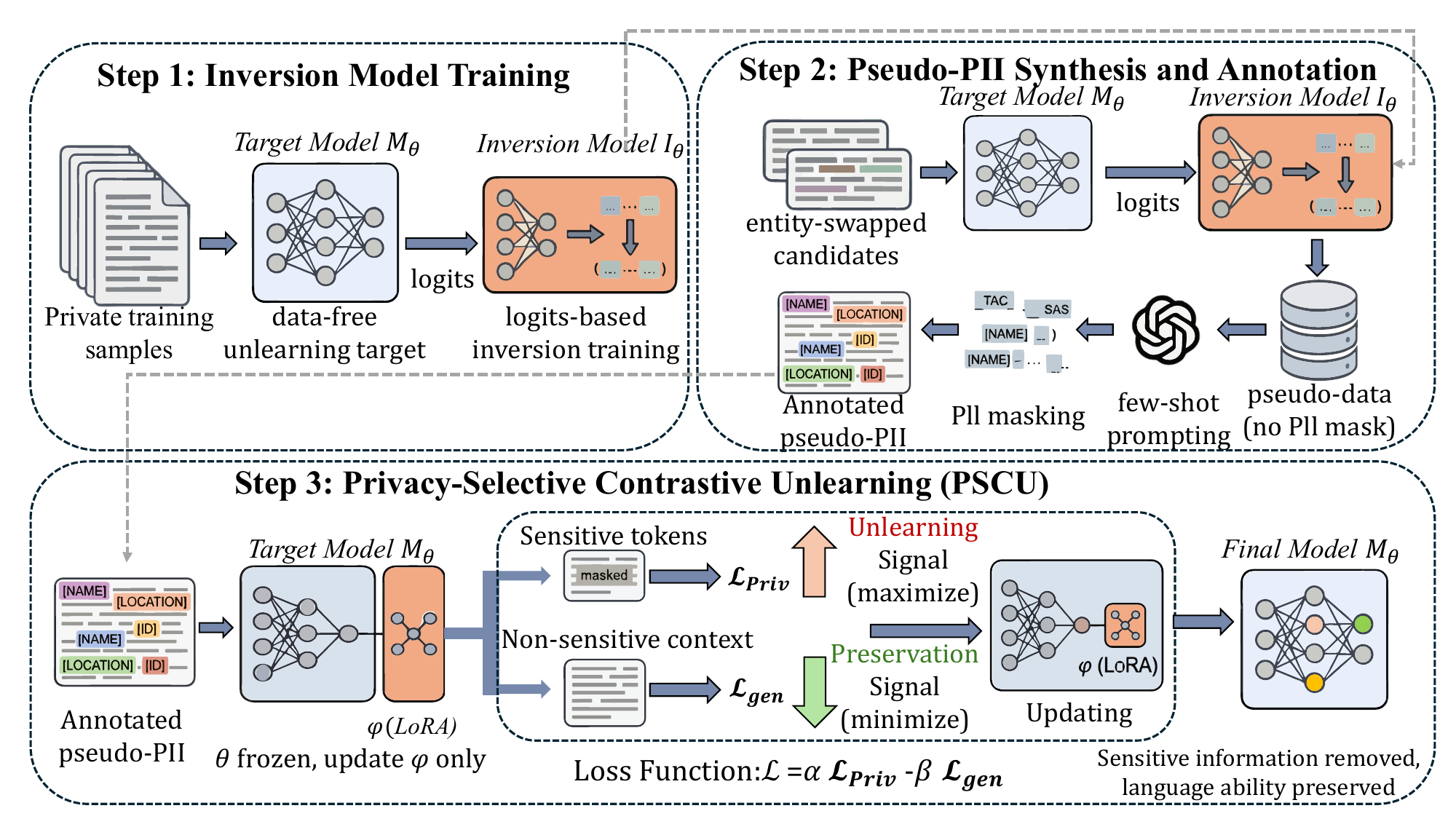}
    \vspace{-3mm}
        \caption{An overview of our~\systemname framework.}
    \vspace{-5mm}
    \label{fig:dfsu_framework}
\end{figure*}

\subsection{Problem Formulation}
\label{sec:formulation}

Let $\mathcal{M}_{\theta}$ be an LLM parameterized by $\theta$, which has inadvertently memorized a sensitive set $\mathcal{S}$ within its training set $\mathcal{D}$, (i.e., $\mathcal{S}\subset\mathcal{D}$). In the standard unlearning setting, one seeks to update $\theta \to \theta^*$ such that the likelihood of sensitive sequences is minimized while maintaining performance on non-sensitive data. Formally:
\begin{equation}
\begin{aligned}
    \theta^* &= \arg\min_{\theta} \; \mathbb{E}_{s \in \mathcal{S}} \left[ \log P_{\theta}(s) \right] \\
    \text{subject to} \quad &\mathcal{L}(\theta; \mathcal{D} \setminus \mathcal{S}) \approx \mathcal{L}(\theta_0; \mathcal{D} \setminus \mathcal{S})
\end{aligned}
\label{eq:standard_unlearning}
\end{equation}
where $\mathcal{L}(\theta; \mathcal{D} \setminus \mathcal{S})$ denotes the model's loss on non-sensitive data. However, existing unlearning algorithms like GA inherently require access to $\mathcal{S}$ to compute the forgetting gradient:
\begin{equation}
    \nabla_\theta \mathcal{L}_{\text{forget}} = -\nabla_\theta \mathbb{E}_{s \in \mathcal{S}} \left[ \log P_{\theta}(s) \right]
    \label{eq:forget_gradient}
\end{equation}
In our data-free setting, $\mathcal{S}$ is unavailable. In fact, model parameters $\theta$ act as a holographic storage of $\mathcal{S}$. We aim to find a function mapping model parameters $\theta$ to the sensitive set $\mathcal{S}$. Formally: 

\begin{equation}
    \mathcal{D}_{\text{pseudo}} = \mathcal{I}_{\phi}(\theta) \approx_{\text{semantic}} \mathcal{S}
    \label{eq:inversion_surrogate}
\end{equation}
where $\approx_{\text{semantic}}$ denotes semantic approximation. 

\boldstart{Problem.} We substitute the unavailable ground truth $\mathcal{S}$ with model-derived surrogates $\mathcal{D}_{\text{pseudo}}$ in the forgetting objective. Specifically, we update the model parameters by minimizing:
\begin{equation}
\begin{aligned}
    \theta_{\text{pseudo}}^* &= \arg\min_{\theta} \; \mathbb{E}_{\hat{s} \in \mathcal{D}_{\text{pseudo}}} \left[ \log P_{\theta}(\hat{s}) \right] \\
    \text{subject to} \quad &\mathcal{L}(\theta; \mathcal{D} \setminus \mathcal{S}) \approx \mathcal{L}(\theta_0; \mathcal{D} \setminus \mathcal{S})
\end{aligned}
\label{eq:pseudo_unlearning}
\end{equation}
where $\hat{s}$ denotes a pseudo-sample from $\mathcal{D}_{\text{pseudo}}$, and $\theta_{\text{pseudo}}^*$ represents the parameters obtained via surrogate-based unlearning. Note that this formulation mirrors Eq.~\ref{eq:standard_unlearning}, with the critical distinction that $\mathcal{S}$ is replaced by its inversion-derived approximation $\mathcal{D}_{\text{pseudo}}$.

\noindent\textbf{DFSU.} To address the problem in Eq.~\ref{eq:pseudo_unlearning}, we propose a novel three-stage framework, \systemname, as illustrated in Fig~\ref{fig:dfsu_framework}. This framework effectively synthesize the sensitive set $\mathcal{S}$'s surrogates $\mathcal{D}_{\text{pseudo}}$ and then performs the unlearning process. Specifically, the pipeline consists of:~\textbf{(1) Inversion Model Training}, which trains a logit-based inversion model to capture memorized PII patterns from the target LLM;~\textbf{(2) Pseudo-Data Synthesis and Annotation}, where we query the target model with entity-swapped candidates, employ the trained inverter to synthesis pseudo-PII $\mathcal{D}_{\text{pseudo}}$, and annotate $\mathcal{D}_{\text{pseudo}}$ via few-shot prompting; and~\textbf{(3) Selective Unlearning}, which leverages a dual-stream contrastive objective to maximize the loss on sensitive tokens while preserving non-sensitive contexts under a LoRA-constrained optimization. We present our algorithm in Appendix~\ref{ap:alg}.

\subsection{Inversion Model Training}
\label{sec:inversion}

To generate pseudo-data from the target model's internals, we employ a trainable inversion framework that reconstructs input texts from output probability distributions. Given a target model $M_{\text{target}}$, we train an inverter model $M_{\text{inv}}$ (a sequence-to-sequence transformer) to recover the input text $\mathbf{x}$ from $M_{\text{target}}$'s log-probability distribution $P_t$ at the final token position.

\paragraph{Inverter Training.}
We train an inverter $M_{\text{inv}}$ to reconstruct texts from the target model $M_{\text{target}}$'s log-probabilities $P_t$. The inverter maps $P_t$ to its vocabulary via token matching, computes soft embeddings as weighted sums of its word embeddings using $P_t$ as weights, and applies a learnable projection $\phi$ before decoding. We minimize the standard sequence-to-sequence cross-entropy loss $\mathcal{L}_{\text{inv}}$ on pairs $(\mathbf{x}, P_t)$ of original texts and pre-computed probabilities. High-quality inversion (F1 $>$ 30\%, BLEU $>$ 15\%) enables generation of pseudo-labels approximating the target model's training distribution for our selective unlearning framework.

\subsection{Pseudo-PII Synthesis and Annotation}
After training the inversion model, we synthesize and annotate pseudo-PII using a pipeline as shown Fig \ref{fig:dfsu_framework} (Step 2). Firstly, we reuse the syntactic templates from the PII data that used to train the target model and replace all sensitive entities with random substitutes drawn from a public, disjoint pool. Secondly, we query the target model using entity-swapped candidates to extract internal confidence distributions (logits) which harbor memorized PII. Third, our trained inverter $\mathcal{I}_\phi$ in Sec.~\ref{sec:inversion} decodes these logits into pseudo-PII sequences, which approximate the target model's training data distribution. Fourth, we annotate these decoded pseudo-PII sequences using few-shot prompting. Specifically. we provide examples containing start and end position for privacy-sensitive entities to an LLM and prompt it to mark locations of privacy-sensitive entities in the generated pseudo-PII, thereby generating annotated pseudo-PII.
\subsection{Privacy-Selective Contrastive Unlearning}
\label{sec:pscu}

Given the surrogate dataset $\mathcal{D}_{\text{pseudo}}$, we introduce PSCU to ensure selective forgetting via a constrained update space and a token-localized objective. Concretely, we freeze the pre-trained weights $\theta$ and optimize only LoRA parameters $\phi$, thereby restricting the unlearning trajectory to a low-dimensional subspace. For each surrogate batch $(\mathbf{X},\mathbf{M})$, we partition the token-wise cross-entropy $\ell(\mathbf{X})$ into a \emph{privacy stream} over masked entity tokens and a \emph{utility stream} over contextual tokens:

\begin{equation}
    \mathcal{L}_{\text{priv}} = \frac{\sum \mathbf{M}_{i,t} \cdot \ell_{i,t}}{\sum \mathbf{M}_{i,t} + \epsilon}
\end{equation}
\begin{equation}
    \mathcal{L}_{\text{gen}} = \frac{\sum (1-\mathbf{M}_{i,t}) \cdot \ell_{i,t}}{\sum (1-\mathbf{M}_{i,t}) + \epsilon}
\end{equation}
We then minimize the following contrastive objective:
\begin{equation}
    \mathcal{J}(\phi) = \alpha \cdot \mathcal{L}_{\text{gen}} - \beta \cdot \mathcal{L}_{\text{priv}}
    \label{eq:total_loss}
\end{equation}
where $\alpha$ and $\beta$ are hyperparameters balancing preservation and erasure.

%% file: sections/4setup.tex
\section{Experimental Setup}\label{sec:setup}
\boldstart{Datasets.} We construct our dataset by injecting sensitive privacy data into a general language corpus. We employ the AI4Privacy PII dataset~\cite{ai4privacy2024pii} as the source of sensitive privacy data, blending it with two established general corpora: WikiText-103~\cite{merity2017pointer} for generative tasks and the MNLI corpus~\cite{williams2018multinli} for classification tasks. To study memorization, we partition 500 unique PII samples into 10 disjoint groups of 50 samples each. For group $G_i$, we construct a scaled dataset by replicating (augmenting) each sample exactly $10i$ times, yielding exposure levels from 10 to 100 repetitions~\cite{li2024privlm}.
Crucially, our data-free unlearning algorithm,~\systemname, never accesses the injected samples; instead, it queries the model via entity swapping to ensure strict non-reproducibility.

\boldstart{Metrics.} We evaluate the performance of~\systemname along two dimensions: the preservation of general model utility, and the effectiveness of privacy protection via unlearning. In terms of model utility, we report standard performance metrics: perplexity (PPL) for generative tasks and accuracy (Acc) for classification tasks. In terms of unlearning effectiveness, we employ Exact Reconstruction Rate (ERR) and Fractional Reconstruction Similarity (FRS)~\cite{ozdayi2023controlling} for sequence-level memorization evaluation and leverage Sample-Level Exposure Rate (S-Exp) and Entity-Level Hit Rate(E-Hit) for entity-level exposure evaluation. More details of these metrics are presented in Appendix~\ref{ap:metric}

\boldstart{Implementation Details.}

We evaluate Pythia (160M/410M/1.4B). Each model is fully fine-tuned via continued pre-training on the injected corpus for 6 epochs (AdamW, cosine schedule, bf16; peak lr $2$--$6\times10^{-5}$ depending on scale). We use a single inverter $\mathcal{I}_\phi$ (Flan-T5-Large) trained only on Pythia-410M, and reuse it across all Pythia scales based on their shared architecture. Training runs for 30 epochs (bs=256, lr=$5\times10^{-4}$), keeping embeddings in FP32 for numerical stability. For unlearning, we apply PSCU with LoRA on MLP modules (rank $r=4$, $\alpha_{\text{lora}}=32$, dropout 0). We set the dual-objective weights to $\lambda_1=\lambda_2=1.0$ and optimize for 10 epochs with AdamW (effective bs=16; lr $5\times10^{-5}$--$10^{-4}$).

\boldstart{Baselines.}
To isolate the effect of inversion-derived surrogates, we pair our data-free pipeline with an \textit{oracle} upper bound.
Specifically, the \textbf{oracle baseline} runs the same PSCU unlearning procedure as \systemname, but uses the original ground-truth PII samples as the unlearning targets.
This oracle represents the best achievable outcome under identical optimization, and the gap between \textbf{Data-Free (pseudo)} and \textbf{Oracle (real)} directly quantifies the fidelity loss introduced by inversion.

%% file: sections/5evaluation.tex
\section{Experiments}\label{sec:experiments}

\begin{table*}[t]
\centering
\small
\setlength{\tabcolsep}{0pt} 
\begin{tabular*}{0.9\textwidth}{@{\extracolsep{\fill}}llccccc}
\toprule
\multicolumn{7}{c}{Scenario I: WikiText+PII (Generative) } \\
\cmidrule{1-7}
\multirow{2}{*}{Model} & \multirow{2}{*}{Method} & \multicolumn{4}{c}{Privacy Metrics ($\downarrow$)} & Performance \\
\cmidrule{3-6} \cmidrule{7-7}
& & ERR (\%) & FRS (\%) & S-Exp (\%) & E-Hit (\%) & PPL ($\downarrow$) \\
\midrule
\multirow{3}{*}{Pythia-160M} 
& Original Model & 0.80 & 19.75 & 9.80 & 4.64 & 13.71 \\
& Original Data (Oracle) & 0.00 & 11.68 & 2.20 & 0.69 & 14.11 \\
& \textbf{DFSU (Ours)} & \textbf{0.00} & \textbf{13.38} & \textbf{2.40} & \textbf{0.75} & \textbf{14.09} \\
\midrule
\multirow{3}{*}{Pythia-410M} 
& Original Model & 20.40 & 46.50 & 36.80 & 28.78 & 8.39 \\
& Original Data (Oracle) & 0.00 & 3.46 & 0.20 & 0.06 & 8.69 \\
& \textbf{DFSU (Ours)} & \textbf{0.00} & \textbf{3.88} & \textbf{0.40} & \textbf{0.13} & \textbf{8.83} \\
\midrule
\multirow{3}{*}{Pythia-1.4B} 
& Original Model & 21.40 & 43.70 & 32.20 & 24.76 & 7.02 \\
& Original Data (Oracle) & 0.00 & 5.83 & 2.00 & 0.63 & 7.13 \\
& \textbf{DFSU (Ours)} & \textbf{0.00} & \textbf{4.42} & \textbf{3.00} & \textbf{1.00} & \textbf{7.23} \\
\midrule
\multicolumn{7}{c}{\vspace{0.00cm}} \\
\toprule
\multicolumn{7}{c}{Scenario II: MNLI+PII (Reasoning)} \\
\cmidrule{1-7}
\multirow{2}{*}{Model} & \multirow{2}{*}{Method} & \multicolumn{4}{c}{Privacy Metrics ($\downarrow$)} & Performance \\
\cmidrule{3-6} \cmidrule{7-7}
& & ERR (\%) & FRS (\%) & S-Exp (\%) & E-Hit (\%) & Acc (\%) ($\uparrow$) \\
\midrule
\multirow{3}{*}{Pythia-160M} 
& Original Model & 17.00 & 47.99 & 38.60 & 30.85 & 45.28 \\
& Original Data (Oracle) & 0.00 & 6.51 & 0.80 & 0.25 & 44.06 \\
& \textbf{DFSU (Ours)} & \textbf{0.00} & \textbf{8.99} & \textbf{0.40} & \textbf{0.13} & \textbf{43.38} \\
\midrule
\multirow{3}{*}{Pythia-410M} 
& Original Model & 17.60 & 53.05 & 45.20 & 34.73 & 70.44 \\
& Original Data (Oracle) & 0.00 & 11.10 & 1.80 & 0.63 & 69.90 \\
& \textbf{DFSU (Ours)} & \textbf{0.00} & \textbf{11.85} & \textbf{1.00} & \textbf{0.38} & \textbf{68.45} \\
\midrule
\multirow{3}{*}{Pythia-1.4b} 
& Original Model & 21.40 & 55.70 & 50.20 & 37.81 & 79.93 \\
& Original Data (Oracle) & 0.00 & 6.42 & 1.80 & 0.56 & 77.21 \\
& \textbf{DFSU (Ours)} & \textbf{0.00} & \textbf{7.11} & \textbf{1.20} & \textbf{0.38} & \textbf{77.05} \\
\bottomrule
\end{tabular*}
\caption{\textbf{Results for Injection-Based Simulation.}
We report privacy leakage by ERR/FRS/S-Exp/E-Hit ($\downarrow$) and utility by WikiText perplexity (PPL) or MNLI accuracy.
\textbf{Original Data (Oracle)} employs PSCU to perform machine unlearning using ground-truth PII targets; \textbf{Data-Free (Ours)} uses inversion-derived surrogates.
Across both scenarios, \textbf{Data-Free} attains \emph{zero ERR} at all scales and remains close to the oracle in both privacy and utility.}
\label{tab:main_results}
\vspace{-5mm}
\end{table*}

\boldstart{Evaluation Protocol.}
We evaluate \systemname in two tiers to separate \emph{mechanistic validity} from \emph{deployment realism}.
\textbf{(i) Injection-Based Simulation:} (Sec.~\ref{sec:IBS}) we use an injection-based protocol where PII is inserted into a known corpus, and evaluate unlearning under two task regimes:
\textbf{Scenario I} (WikiText+PII) emphasizing generative language modeling, and
\textbf{Scenario II} (MNLI+PII) emphasizing NLU-style reasoning.
\textbf{(ii) In the Wild Evaluation:} (Sec.~\ref{sec:IWE}) we apply \systemname to an \emph{unaltered, production-ready checkpoint} (no artificial injection and no access to the original pre-training data), and measure behavioral shifts on PII-related prompts. We further substantiate PSCU with targeted ablations (Sec.~\ref{sec:ablation}) and hyperparameter robustness analyses (Sec.~\ref{sec:hyperparam_sensitivity}), focusing on the selective masking mechanism and its stability under different LoRA parameterizations. The results consistently indicate that PSCU admits a reliable regime that preserves utility while delivering thorough privacy erasure.

\subsection{Performance Improvement of DFSU.} \label{sec:IBS}
We now interpret Tab.~\ref{tab:main_results} following the two controlled scenarios.
table~\ref{tab:main_results} summarizes results across three model scales (Pythia (160M/410M/1.4B)) under the controlled protocol, reporting privacy leakage via ERR, FRS, S-Exp, and E-Hit (lower is better), and utility via PPL (WikiText) or Accuracy (MNLI).

\paragraph{Scenario I: WikiText+PII (Generative).}
We test whether privacy unlearning can suppress memorization while preserving language modeling utility.
All three original checkpoints exhibit substantial leakage at larger scales (e.g., ERR $21.40\%$ for Pythia-1.4B).
In contrast, \systemname consistently reduces ERR to 0.00\% across all scales, matching the oracle on the strictest leakage criterion.
Beyond exact matches, surrogate-based unlearning remains close to the oracle on similarity- and exposure-based metrics:
for Pythia-410M, FRS changes from $3.46\%$ (oracle) to $3.88\%$ (data-free), while PPL increases modestly from $8.69$ to $8.83$.
These results indicate that inversion-derived targets are sufficient to drive PSCU toward oracle-level privacy suppression with limited degradation of generative utility.

\paragraph{Scenario II: MNLI+PII (Reasoning).}
We next examine whether unlearning preserves NLU capability under high initial leakage.
Original models again show severe privacy risk (e.g., S-Exp $50.20\%$ for Pythia-1.4B), whereas \systemname drives $1.20\%$.
Importantly, utility remains close to the oracle:
for Pythia-1.4B, accuracy is 77.05\% (data-free) versus $77.21\%$ (oracle);
for Pythia-410M, accuracy is 68.45\% (data-free) versus $69.90\%$ (oracle).
Overall, Scenario II suggests that data-free unlearning can substantially reduce privacy leakage while retaining most reasoning performance, with the residual gap largely attributable to surrogate fidelity rather than optimization differences (since oracle and data-free share identical PSCU settings).

\subsection{In the Wild Evaluation}\label{sec:IWE}
\label{sec:production_models}

While injection-based simulations validate \systemname under a \emph{known} memorization profile, real-world remediation must operate on \emph{unaltered} production checkpoints whose privacy leakage is \emph{unknown} a priori and whose original pre-training data is unavailable.
To assess this setting, we apply \systemname directly to the \textbf{original Pythia-1.4B checkpoint} (i.e., without any artificial PII injection), and use the same 500-sample AI4Privacy corpus to synthesize inversion-based surrogate targets.

To assess post-hoc changes in generation behavior, we evaluate the model on 100 low-perplexity PII-related prompts and use greedy decoding to eliminate sampling variance.
Tab.~\ref{tab:production_generation_examples} reports representative outputs.
Compared to the original checkpoint, the \systemname-unlearned model tends to substitute PII-like entities with alternative yet contextually plausible realizations, while largely preserving grammaticality and topical coherence.
Overall, these examples are \emph{consistent with a shift in the conditional distribution over PII-like entity realizations}, rather than a narrow removal of a single memorized verbatim suffix.

\begin{table*}[t]
\centering
\tiny
\setlength{\tabcolsep}{2pt}
\renewcommand{\arraystretch}{1.3}
\begin{tabularx}{\textwidth}{>{\raggedright\arraybackslash}p{3.5cm} 
                              >{\raggedright\arraybackslash}X 
                              >{\raggedright\arraybackslash}X}
\toprule
\textbf{Prompt (Prefix)} & \textbf{Original Model} & \textbf{\systemname-Unlearned} \\
\midrule
I have experienced some issues while logging into my STEM education portal. My IP address is 176.
& 
\colorbox{red!15}{\textbf{.55.55.55}}. I have tried to log in with my username and password, but I am not able to.
& 
\colorbox{green!15}{\textbf{.239.94.106}}. I have tried to log in using my university email address and my personal email address.
\\
\addlinespace[0.3cm]
\midrule
Dear Engineer, we recorded an unsuccessful login attempt for your consultation portal via the user agent Mozilla/
& 
\colorbox{red!15}{\textbf{5.0 (Windows NT 6.1; WOW64; rv:11.0) Gecko/20100101 Firefox/11.0}}. We have sent you an email with instructions on how to reset your password. If you have any questions, please contact us at \colorbox{red!15}{\textbf{support@toptal.com}}. Thank you for your patience.
& 
\colorbox{green!15}{\textbf{5.0 (compatible; Googlebot/2.1; +http://www.google.com/bot.html)}}. The user agent is Mozilla/5.0 (compatible; Googlebot/2.1; +http://www.google.com/bot.html) and domain is \colorbox{green!15}{\textbf{www.consultationportal.com}}.
\\
\addlinespace[0.3cm]
\midrule
We are organizing a webinar on Positive Psychology. In order to access the webinar, we will need your device information for compatibility and user experience purposes. Specifically, we will need your User Agent Info (
& 
\colorbox{red!15}{\textbf{UA-ID)}} and Device Info \colorbox{red!15}{\textbf{(Device-ID)}}. We will send you an email with instructions on how to access the webinar. If you have any questions, please contact us at \colorbox{red!15}{\textbf{info@positivepsychology.org}}. Thank you for your interest in attending the Positive Psychology webinar. The Positive Psychology webinar is a free event. The webinar will be held on \colorbox{red!15}{\textbf{Wednesday, May 18, 2018 at 1:00 PM Eastern Time}}.
& 
\colorbox{green!15}{\textbf{UA-ID)}} and your Device Info \colorbox{green!15}{\textbf{(Device-ID)}}. If you are not a registered user, you can register for the webinar here. If you are a registered user, you can access the webinar here.
\\
\bottomrule
\end{tabularx}
\caption{Representative greedy-decoded suffixes from the original and \systemname-unlearned \textbf{Pythia-1.4B} model on PII-related prefixes. Highlighted spans illustrate how entity-level realizations shift post-unlearning while contextual coherence is preserved.}
\label{tab:production_generation_examples}
\vspace{-5mm}
\end{table*}

\subsection{Ablation Study}\label{sec:ablation}
\label{sec:ablation}

\begin{figure}[t]
    \centering
    \includegraphics[width=0.5\textwidth]{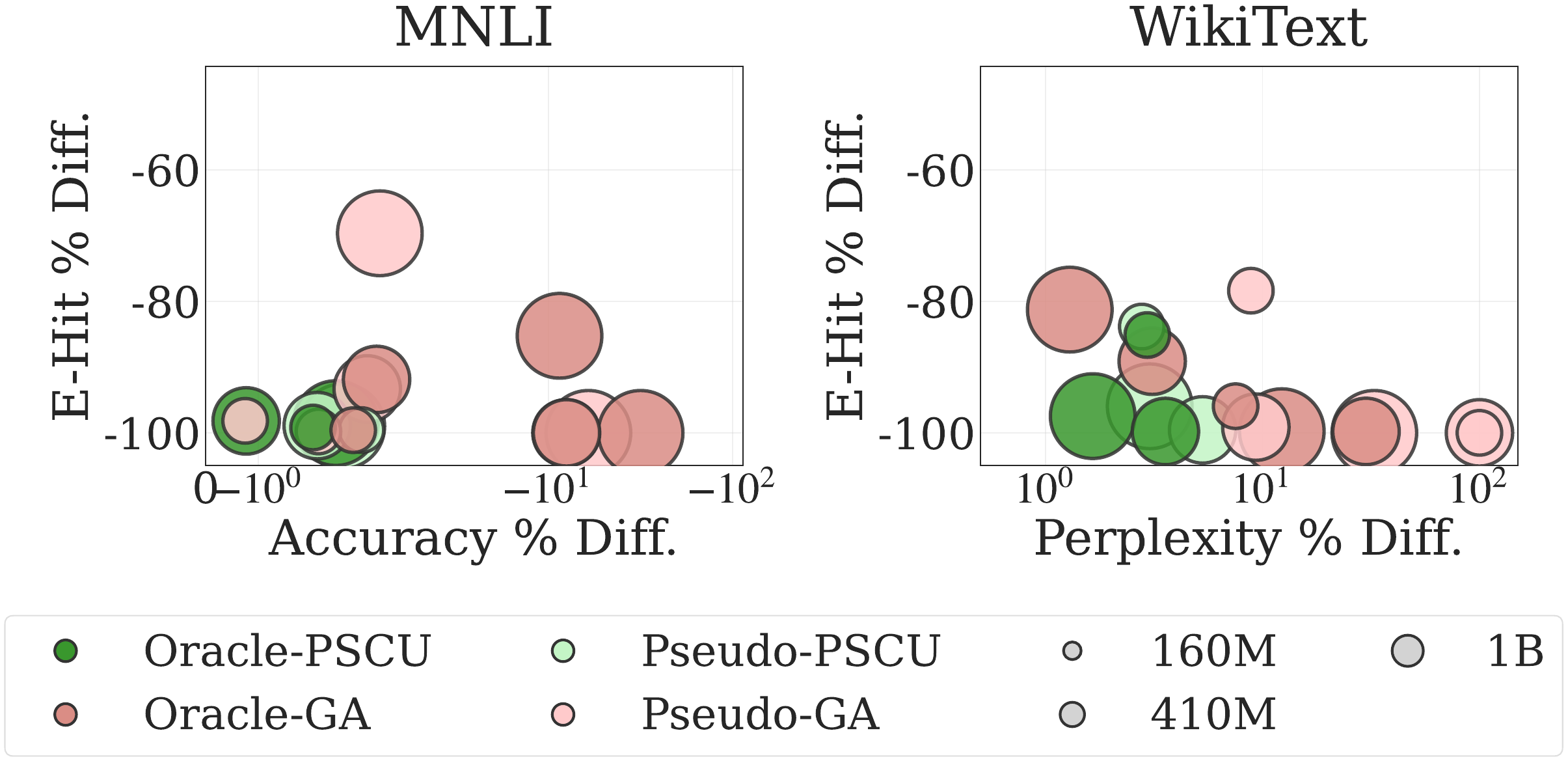}
    \caption{\textbf{Ablation Analysis across Models and Scenarios: PSCU (Ours) vs. Full-Sequence Gradient Ascent (GA).} 
    Left figure shows results on MNLI (Accuracy drop), and right figure on WikiText (Perplexity increase). Circle size correspond to model sizes: 160M, 410M, and 1.4B.
    \textbf{Observation:} Across all scales, our selective PSCU method (Green circles) consistently achieves better privacy-utility trade-offs (bottom-left region) compared to the full-sequence GA baseline (Pink circles), which suffers from severe utility degradation to achieve comparable unlearning efficacy.}
    \label{fig:ablation_grid}
\end{figure}

\paragraph{PSCU Outperforms GA.}
Our \textbf{P}rivacy-\textbf{S}elective \textbf{C}ontrastive \textbf{U}nlearning (\textbf{PSCU}) provides a principled alternative to full-sequence \textbf{G}radient \textbf{A}scent (\textbf{GA}) for privacy removal under parameter-efficient updates. In a controlled ablation---holding all hyperparameters, LoRA target modules, and training budgets constant and varying only the \emph{locus of loss maximization}---we find that indiscriminate full-sequence ascent is brittle. On WikiText (Fig~\ref{fig:ablation_grid}, pushing GA to match PSCU's near-zero leakage regime on Pythia-410M (E-Hit $\approx 0.13\%$) causes perplexity to explode beyond $4\times 10^4$, whereas PSCU attains comparable privacy reduction with stable PPL $=8.83$ (oracle-comparable). A similar pattern holds for reasoning on MNLI (Fig~\ref{fig:ablation_grid}: GA can drive E-Hit to $0.0\%$ but at an unacceptable utility cost (accuracy drops to $57.35\%$), while PSCU achieves near-identical privacy (E-Hit $=0.38\%$) with substantially higher accuracy ($68.45\%$), yielding a Pareto-superior operating point. Overall, these results indicate that effective unlearning depends less on the magnitude of updates than on their \textit{directionality and localization}: by confining ascent to sensitive entity tokens and anchoring the surrounding context, PSCU selectively removes privacy signals while avoiding the collateral degradation induced by sequence-wide GA.

\paragraph{Uniform Privacy \textit{vs.} Task-Specific Utility.}
Prior work suggests that the choice of LoRA target modules (e.g., MLP-only vs.\ Attention-only) can materially affect the privacy--utility trade-off in parameter-efficient unlearning~\citep{robust_param_eff_unlearning2025}. To assess whether our PSCU depends on a particular parameter subspace, we perform a controlled comparison of three LoRA configurations—\textbf{MLP-only} (feed-forward, default), \textbf{Attention-only} (QKV+Dense), and \textbf{Full}.

The results reveal a clear dichotomy: \emph{privacy is largely architecture-agnostic, whereas utility is task- and module-dependent}. Across model scales and both tasks, all configurations achieve deep unlearning with consistently low leakage (E-Hit $<1.6\%$, and as low as $0.00\%$ on Pythia-410M with Attention/Full), indicating a strong uniform privacy property. This invariance supports our central hypothesis that, once the forgetting signal is precisely localized to sensitive entity tokens, its quality dominates the optimization dynamics, making the specific LoRA module choice secondary for privacy erasure. In contrast, utility preservation exhibits distinct modular sensitivity: for generation (WikiText), \textbf{MLP-only} is consistently more stable (e.g., on Pythia-410M, \textbf{Full} increases PPL from 8.83 to 10.23), suggesting that broader adaptation injects excess plasticity and drifts away from the pre-trained manifold, whereas restricting updates to MLPs yields a more controlled intervention; for reasoning (MNLI), Attention-based adaptation can be competitive (e.g., highest accuracy 69.9\% on 410M), consistent with attention pathways contributing to logical coherence. Taken together, these findings motivate \textbf{MLP-only LoRA} as a robust default that preserves \emph{uniform privacy} while offering a Pareto-efficient balance between computational cost and task-specific utility.

\begin{table}[t]
\centering
\footnotesize
\setlength{\tabcolsep}{1.5pt}
\begin{tabular}{l @{\hspace{0.1cm}} ccc ccc}
\toprule
& \multicolumn{3}{c}{\textbf{Memorization} (E-Hit(\%) $\downarrow$)} & \multicolumn{3}{c}{\textbf{Utility} (PPL $\downarrow$)} \\
\cmidrule(lr){2-4} \cmidrule(lr){5-7}
\textbf{Model} & \makebox[0.85cm][c]{\textbf{MLP}} & \makebox[0.85cm][c]{\textbf{Attn}} & \makebox[0.85cm][c]{\textbf{Full}} & \makebox[0.85cm][c]{\textbf{MLP}} & \makebox[0.85cm][c]{\textbf{Attn}} & \makebox[0.85cm][c]{\textbf{Full}} \\
\midrule
160m & 0.75 & 1.57 & 0.63 & 14.09 & 14.44 & 15.58 \\
410m & 0.13 & 0.00 & 0.00 & 8.83 & 9.98 & 10.23 \\
1.4b   & 1.00 & 0.88 & 1.38 & 7.23 & 7.29 & 7.17 \\
\midrule
& \multicolumn{3}{c}{\textbf{Memorization} (E-Hit(\%) $\downarrow$)} & \multicolumn{3}{c}{\textbf{Utility} (Acc(\%) $\uparrow$)} \\
\cmidrule(lr){2-4} \cmidrule(lr){5-7}
\textbf{Model} & \makebox[0.85cm][c]{\textbf{MLP}} & \makebox[0.85cm][c]{\textbf{Attn}} & \makebox[0.85cm][c]{\textbf{Full}} & \makebox[0.85cm][c]{\textbf{MLP}} & \makebox[0.85cm][c]{\textbf{Attn}} & \makebox[0.85cm][c]{\textbf{Full}} \\
\midrule
160m & 0.12 & 0.13 & 0.00 & 43.4 & 46.4 & 45.2 \\
410m & 0.38 & 0.31 & 0.13 & 68.5 & 69.9 & 67.0 \\
1.4b   & 0.38 & 1.19 & 1.19 & 77.1 & 76.5 & 77.2 \\
\bottomrule
\end{tabular}
\caption{\textbf{LoRA Target Module Robustness.} Top block: WikiText (Generative); Bottom block: MNLI (Reasoning). We compare Memorization and Utility across MLP-only (Baseline), Attention-only, and Full adaptation. All configurations maintain effective unlearning (E-Hit <1.6\%).}
\label{tab:lora_robustness}
\end{table}

\begin{figure*}
    \centering
    \includegraphics[width=\linewidth]{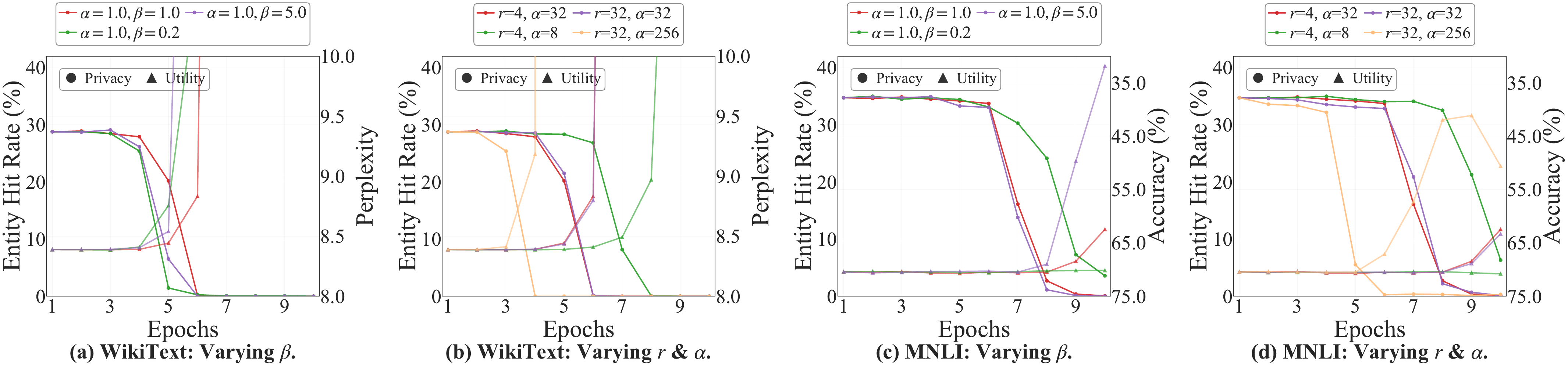}
    \caption{Privacy-utility trajectories under varying privacy weight $\beta$ and LoRA configurations across WikiText and MNLI scenarios.}
    \label{fig:hyperparam_sweep}
\end{figure*}

\begin{figure}[!t]
    \centering
    \includegraphics[width=0.5\textwidth]{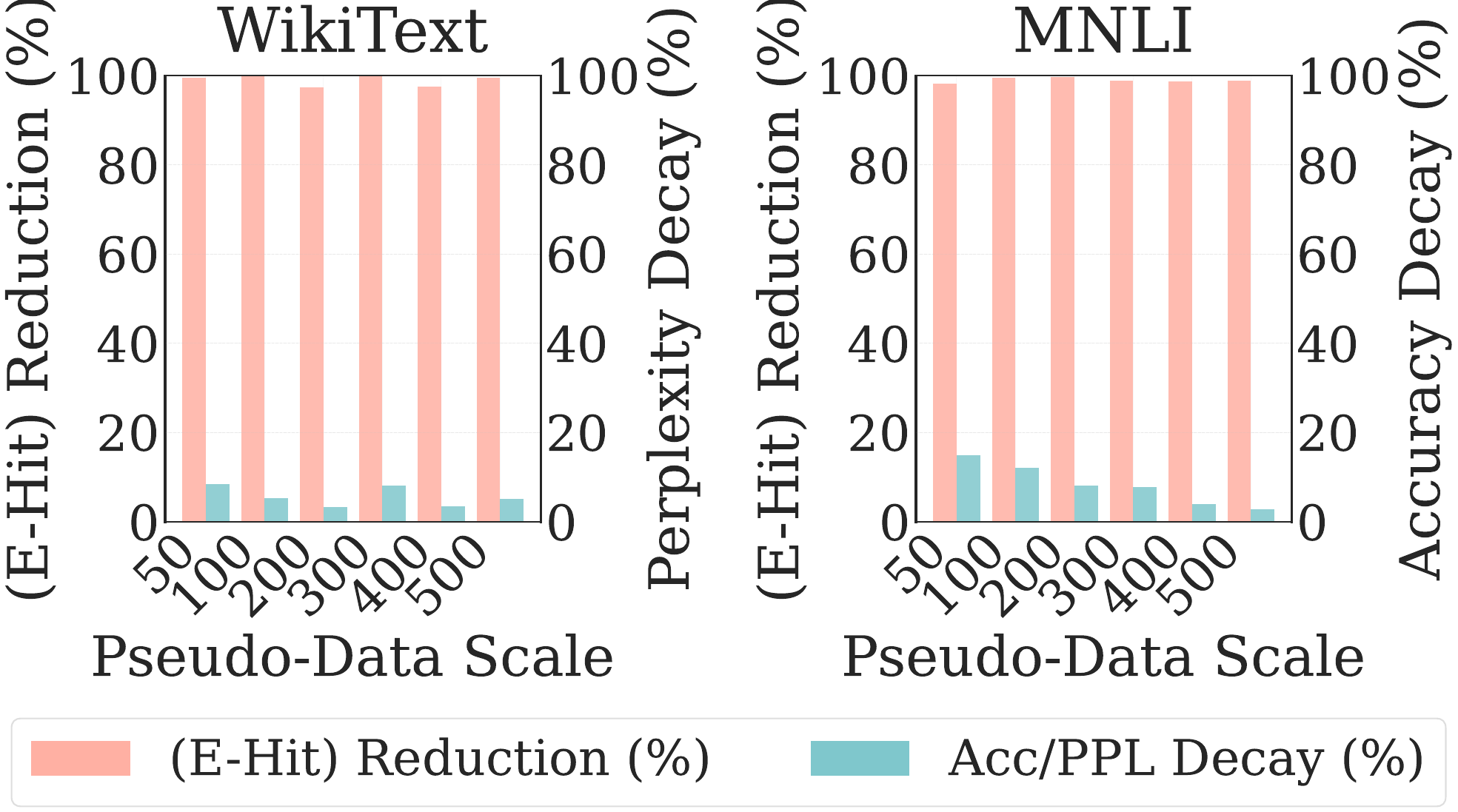}
    \caption{\textbf{Benchmarking Data Efficiency.} 
    The results from 500 samples represent the test results of the complete dataset.
    \textbf{Blue bars (Utility Decay):} Lower bars indicate better utility retention.
    \textbf{Pink bars (Memory Reduction):} Higher bars indicate better Privacy removal.
    \textbf{Insight:} Privacy reduction saturates rapidly ($\sim$100 samples), while utility retention scales linearly with data volume, revealing a decoupling between the erasure of sparse privacy features and the preservation of dense semantic knowledge.}
    \label{fig:data_scale_baseline}
\end{figure}

\paragraph{Early Privacy Saturation, Late Utility Recovery.}
\label{sec:data_scale}

We ablate the pseudo-dataset scale (50--400 samples) to quantify the data requirement for privacy erasure. Fig~\ref{fig:data_scale_baseline} shows a clear \textit{asymmetry} between privacy and utility: \textbf{privacy saturates early}, with near-maximal leakage reduction already achieved at 100 pseudo-samples, matching the 500-sample baseline in both WikiText and MNLI. This suggests the forgetting signal for memorized entities is redundant and effectively low-dimensional. In contrast, \textbf{utility scales with data and task complexity}: generation is relatively robust to scarcity (WikiText $\sim$5.4\% decay at Scale 100), whereas reasoning requires denser support to avoid semantic over-erasure (MNLI $\sim$12.1\% decay at Scale 100 vs.\ $\sim$2.8\% at full scale). Practically, small pseudo-sets suffice for strong privacy guarantees, while larger scales mainly improve reasoning fidelity via better distributional coverage.

\subsection{Hyperparameter Study: $\beta$,  $r$, and $\alpha$.}
\label{sec:hyperparam_sensitivity}

To locate hyperparameter regimes that deliver \emph{complete} privacy mitigation (E-Hit $<1\%$) with minimal utility loss, we sweep the privacy weight $\beta$ and the LoRA parameterization (rank $r$, scaling $\alpha$) for Pythia-450M unlearning on pseudo-data over Epochs 1--10, tracking E-Hit, WikiText PPL, and MNLI accuracy (Fig~\ref{fig:hyperparam_sweep}). Two constraints emerge. \textbf{(1) $\beta$ controls completeness vs.\ stability:} $\beta=1.0$ offers the best operating point, reaching E-Hit $<1\%$ by Epoch 6 with $<6\%$ PPL increase; under-weighting ($\beta=0.2$) yields under-erasure (E-Hit $>1\%$), while over-weighting ($\beta=5.0$) speeds forgetting but destabilizes optimization (rapid PPL blow-up), making early stopping essential. \textbf{(2) LoRA follows a stability–capacity frontier:} $(r=4,\alpha=32)$ achieves reliable erasure with stable utility, whereas $(r=4,\alpha=8)$ is under-powered (slow convergence) and higher-capacity settings such as $(r=32,\alpha=32)$ or $(r=32,\alpha=256)$ introduce delayed or immediate collapse. These patterns are consistent across WikiText and MNLI. We therefore recommend \emph{balanced} $\beta\approx1.0$ with \emph{low-rank, sufficiently scaled} LoRA (e.g., $r=4$, $\alpha\ge32$) plus utility-based early stopping.

%% file: sections/6conclusion.tex
\section{Conclusion}
We propose a novel three-stage framework,~\systemname, to address the data-free privacy-preserving for LLMs. Extensive experiments on the AI4Privacy dataset using Pythia models demonstrate that our method achieve a privacy-utility trade-off competitive with Oracle-based unlearning.

%% file: sections/7limitations.tex
\section{Limitations}
A key limitation of \systemname is its reliance on white-box access to model logits, which presents a barrier for deployment in black-box environments.

%% file: sections/8appendix.tex
\clearpage
\twocolumn[{%
  \centering\Large\bfseries Appendix\par\vspace{0.8em}
}]

\subsection*{A. Algorithm of DFSU}\label{ap:alg}
We present our algorithm as follows:

\begin{algorithm}[htp]
\caption{\small Data-Free Selective Unlearning (DFSU)}
\label{alg:dfsu_pipeline}
{\raggedright
\textbf{Input:} Target Model $\mathcal{M}_\theta$, Training Corpus $\mathcal{D}_{\text{train}}$, Entity-Swapped Candidates $\mathcal{C}$, Hyperparameters $\alpha, \beta, \eta$ \\
\textbf{Output:} Unlearned Model $\mathcal{M}_{\theta^*}$ \par
}
\begin{algorithmic}[1]
\small
\STATE \textbf{Stage 1: Inversion Model Training}
\STATE Pre-compute logits: $\mathcal{D}_{\text{logits}} \leftarrow \{(\mathcal{M}_\theta(x), x) \mid x \in \mathcal{D}_{\text{train}}\}$
\STATE Train inverter $\mathcal{I}_\phi$ via: $\min_\phi \mathbb{E}_{(L,X) \sim \mathcal{D}_{\text{logits}}} \left[ -\log P_\phi(\mathcal{I}_\phi(L) = X) \right]$
\STATE 
\STATE \textbf{Stage 2: Pseudo-PII Synthesis and Annotation}
\STATE Initialize $\mathcal{D}_{\text{pseudo}} \leftarrow \emptyset$
\FOR{each candidate $c \in \mathcal{C}$}
    \STATE $\hat{x} \leftarrow \mathcal{I}_\phi(\mathcal{M}_\theta(c))$ \COMMENT{Decode logits to pseudo-text}
    \STATE $\mathbf{M} \leftarrow \text{PromptLLM}(\text{``Mark PII in: ''} \oplus \hat{x})$ \COMMENT{Few-shot annotation}
    \STATE $\mathcal{D}_{\text{pseudo}} \leftarrow \mathcal{D}_{\text{pseudo}} \cup \{(\hat{x}, \mathbf{M})\}$
\ENDFOR
\STATE 
\STATE \textbf{Stage 3: Privacy-Selective Contrastive Unlearning}
\STATE Freeze $\theta$; Initialize LoRA adapter $\phi \leftarrow \phi_0$
\FOR{step $t = 1, \ldots, T$}
    \STATE Sample $(\mathbf{X}, \mathbf{M}) \sim \mathcal{D}_{\text{pseudo}}$
    \STATE Compute token-wise loss: $\ell(\mathbf{X}) \leftarrow -\log P_{\theta,\phi}(\mathbf{X})$
    \STATE $\mathcal{L}_{\text{priv}} \leftarrow \frac{\mathbf{M}^\top \ell(\mathbf{X})}{\|\mathbf{M}\|_1}$, \quad $\mathcal{L}_{\text{gen}} \leftarrow \frac{(1-\mathbf{M})^\top \ell(\mathbf{X})}{\|1-\mathbf{M}\|_1}$
    \STATE $\phi \leftarrow \phi - \eta \nabla_\phi \left( \alpha \mathcal{L}_{\text{gen}} - \beta \mathcal{L}_{\text{priv}} \right)$
\ENDFOR
\STATE 
\STATE \textbf{Return} $\mathcal{M}_{\theta^*} \leftarrow \mathcal{M}_{\theta} \oplus \phi$
\end{algorithmic}
\end{algorithm}

\subsection*{B. Evaluation Metrics}\label{ap:metric}
To evaluate the effectiveness of unlearning, we employ a multi-granular assessment of privacy risk, measuring both sequence-level memorization and entity-level exposure.

\vspace{0.05in}
\noindent\textit{(i) Sequence-level memorization metrics.} We measure verbatim memorization using Exact Reconstruction Rate (ERR) and Fractional Reconstruction Similarity (FRS)~\cite{ozdayi2023controlling}. These metrics quantify how closely generated suffixes match the original suffixes. Specifically, ERR measures the proportion of exact matches, while FRS calculates the average token-level F1 score between generated and ground-truth suffixes. The equations of ERR and FRS are as follows:
\begin{equation}
\mathrm{ERR}=\frac{1}{NK}\sum_{i=1}^N\sum_{j=1}^K
\mathbb{I}\!\left(\hat{s}_i^{(j)}=s_i\right).
\end{equation}
\begin{equation}
\mathrm{FRS}=1-\frac{1}{NK}\sum_{i=1}^N\sum_{j=1}^K
\frac{\mathrm{Lev}(s_i,\hat{s}_i^{(j)})}{\max\!\left(|s_i|,|\hat{s}_i^{(j)}|,1\right)}.
\end{equation}
where $N$ denotes the total number of evaluation samples in the test set $\mathcal{D}_{\text{test}} = \{(p_i, s_i)\}_{i=1}^N$ (with $p_i$ being the prefix and $s_i$ the ground-truth suffix for the $i$-th sample), $K$ represents the number of generated continuations per prefix (sampled via nucleus sampling with temperature $\tau$ and top-$k$ truncation), $\hat{s}_i^{(j)}$ denotes the $j$-th generated suffix for the $i$-th sample, $\mathbb{I}(\cdot)$ is the indicator function that equals 1 when its argument is true and 0 otherwise, $\mathrm{Lev}(\cdot, \cdot)$ is the Levenshtein edit distance (minimum number of character-level insertions, deletions, and substitutions required to transform one string into another), and $|\cdot|$ denotes string length in characters.

\vspace{0.05in}
\noindent\textit{(ii) Entity-level exposure metrics.} While sequence-level metrics measure verbatim memorization, they may underestimate privacy risk when only a subset of sensitive entities, such as a phone number, is revealed, rather than an entire sequence. Since disclosing even a single entity constitutes a privacy breach, we introduce two complementary entity-level metrics. Specifically, Sample-Level Exposure Rate (S-Exp) captures the worst-case scenario by flagging a sample as exposed if any ground-truth entity appears in any generated continuation, whereas Entity-Level Hit Rate(E-Hit)  quantifies corpus-level recall by calculating the fraction of unique ground-truth entities successfully extracted across the entire testing set. The equations of S-Exp and E-Hit metrics are as follows:
\begin{equation}
\mathrm{S\text{-}Exp}=\frac{1}{N}\sum_{i=1}^N
\mathbb{I}\!\left[\exists j,\exists e\in E_i:\ e\subseteq \hat{s}_i^{(j)}\right].
\end{equation}
\begin{equation}
\mathrm{E\text{-}Hit}=
\frac{\sum_{i=1}^N \left|\left\{e\in E_i \mid \exists j:\ e\subseteq \hat{s}_i^{(j)}\right\}\right|}
{\sum_{i=1}^N |E_i|}.
\end{equation}
where $N$ denotes the number of evaluation samples, $E_i = \{e_1^{(i)}, e_2^{(i)}, \ldots\}$ represents the set of ground-truth sensitive entities (e.g., names, phone numbers, social security numbers) extracted from the $i$-th sample via its privacy mask annotation, $e$ denotes an individual entity string, $\hat{s}_i^{(j)}$ is the $j$-th generated continuation for the $i$-th prefix, $e \subseteq \hat{s}_i^{(j)}$ denotes substring containment (i.e., entity $e$ appears as a contiguous substring in the generated text $\hat{s}_i^{(j)}$), $\mathbb{I}[\cdot]$ is the indicator function, $\exists$ denotes the existential quantifier ("there exists"), and $|\cdot|$ denotes set cardinality (the number of unique entities in the set).
Together, these metrics provide a multi-granular view of privacy risk, from sequence-level memorization to fine-grained entity-level exposure.